\documentclass[aps,prl,twocolumn,showpacs,floatfix,superscriptaddress]{revtex4-1}
\usepackage{graphicx}
\usepackage{amsfonts,amsmath,color}
\usepackage{multirow}
\usepackage{bigstrut}
\usepackage[normalem]{ulem}
\usepackage{color}
\definecolor{dgreen}{rgb}{0,0.7,0}
\def\redw#1{{\color{black} #1}}

\def\la{\langle}
\def\ra{\rangle}
\begin{document}
\title{Non-equilibrium dynamics of  the piston in the Szilard engine}
\author{Deepak Bhat}
\email{deepak.bhat@santafe.edu}
\affiliation{Santa Fe Institute, 1399 Hyde Park Rd., Santa Fe, New Mexico 87501, USA}
\author{Abhishek Dhar}
\affiliation{International Centre for Theoretical Sciences,  Bengaluru 560089, India}
\author{Anupam Kundu}
\affiliation{International Centre for Theoretical Sciences,  Bengaluru 560089, India}
\author{Sanjib Sabhapandit}
\affiliation{Raman Research Institute, Bengaluru 560080, India}

\date{\today}

\begin{abstract}
We consider a Szilard engine in one dimension, consisting of a single particle of mass $m$, moving between a piston of mass $M$, and a heat reservoir at temperature $T$. In addition to an external force, the piston experiences repeated elastic collisions with the particle. We find that the motion of a heavy piston ($M \gg m$), can be described effectively by a Langevin equation.  Various numerical  evidences suggest that the frictional coefficient in the Langevin equation is given by $\gamma = (1/X)\sqrt{8 \pi m k_BT}$, where   $X$ is the position of the piston measured from the thermal wall. Starting from the  exact master equation for the full system and  using a perturbation expansion in $\epsilon= \sqrt{m/M}$,  we integrate out the  degrees of freedom of the particle to obtain the effective Fokker-Planck equation for the piston albeit with a  different frictional coefficient. Our microscopic study shows that the piston is never in equilibrium during the expansion step, contrary to the  assumption made in the usual Szilard engine analysis --- nevertheless the conclusions of Szilard remain valid.
\end{abstract}

\maketitle 

 
\emph{Introduction}.-- The Szilard engine, a simple realization of the Maxwell demon,  is a paradigmatic model  designed to address the conceptual foundations of the second law of thermodynamics~\cite{Szilard}. In apparent violation of the second law, this model envisages a thought experiment, where the system working in  a cycle extracts work from a single heat reservoir by using the information about the initial state of the system. Recently there has been renewed interest in this problem  due to important developments in the areas of stochastic thermodynamics of small systems~\cite{Bustamante} and fluctuation theorems~\cite{Jarzynski2,Jarzynski,Evans,Seifert,Kurchan}. Moreover, recent developments of technology has made it possible to realize this thought experiment in the laboratory~\cite{Toyabe,Koski,Berut}.

In spite of the importance of the Szilard problem,  surprisingly  there have  been very few microscopic studies~\cite{Hatano,Hondou} of its  dynamics  in the original set-up.  In Szilard's analysis, the piston is assumed to be ideal, having infinite mass and its motion is then described by a quasi-static deterministic process. However a realistic  piston has a large but finite mass.   As a result, its motion is strongly affected by fluctuations that we expect to be important in   small systems. Hence it is crucial to understand the stochastic dynamics of the piston.  This is the main aim of this Letter.

\begin{figure}
\centering
\includegraphics[width=2cm,angle=-90]{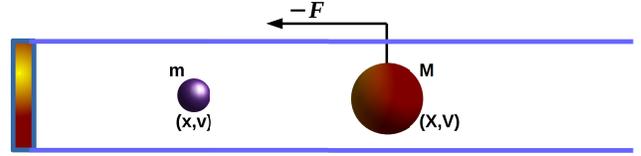}
 \caption{\label{fig1} (Color online) A schematic of the Szilard set-up  in one dimension. A particle of mass m moves ballistically between a thermal reservoir at the left end and another particle (the piston) of mass $M \gg m$ on the right. The piston is subject to an external force $-F$.}\label{f1}
\end{figure}

The basic model consists of a single hard point particle of mass $m$ confined to move in one dimension, between a   piston and a heat reservoir at temperature $T$ (see Fig.~\ref{f1}). The piston itself is taken to be another hard point particle of mass $M>>m$.  Let $x$ and $v$  ($X$ and $V$) respectively be the position and velocity of the particle (piston). On  collisions with the thermal wall at $x=0$, the particle emerges with a velocity  $v>0$, chosen independently at each time from the Rayleigh distribution $f(v)=\beta m v e^{-\beta m v^2/2}$, where $\beta=(k_BT)^{-1}$. The collision between the particle and the piston is taken to be elastic. In between collisions with the wall and the piston, the particle moves ballistically. The piston, apart from collisions with the particle, also experiences an external force $-\mathcal{U}'(X)$.   This dynamics  takes the system to the Gibb's equilibrium state $P_\text{eq}=Z^{-1}\exp{[ -\beta (mv^2/2+M V^2/2 + \mathcal{U}(X))]} \theta(x) \theta(X-x)$, where $Z$ is the partition function.  However the dynamics of the relaxation process is non-trivial, as can be seen even when the piston is held fixed \cite{Bhat}.   

Our set-up is similar to that of the well-known  adiabatic piston problem, where one usually considers the deterministic motion of a heavy piston in presence of a gas of thermodynamically large number of small particles with \cite{Neishtadt,Gruber,Mansour,Cerino,Hondou,Proesmans,Baule,Hoppenau,Chernov,Chernov2,Lebowitz} or without reservoirs \cite{Sinai,Wright}. Here we look at the stochastic dynamics of the piston in the presence of a single particle gas, as required in the Szilard set-up, where fluctuations play an important role.

Our main finding is that, in the limit of heavy piston mass  ($M>>m$), the  effective stochastic dynamics is given by the Langevin equation
\begin{eqnarray}
M\frac{dV}{dt} = -\mathcal{U}'(X)+\frac{k_BT}{X}-\gamma(X) V +\sqrt{2\gamma(X) k_BT}~\eta(t),~ ~~~\label{eom}
\end{eqnarray}
where $\eta(t)$ is Gaussian white noise with $\langle \eta(t)\rangle=0$ and $\langle\eta(t) \eta(t')\rangle=\delta(t-t')$, and the space-dependent dissipation is given by 
\begin{eqnarray}
\gamma(X)= \frac{1}{X}\sqrt{8\pi mk_BT}=:\gamma_{\rm sp}.\label{gamma}
\end{eqnarray}
The second term on the  rhs of Eq.~\eqref{eom} is the pressure term while the frictional and noise terms satisfy the fluctuation-dissipation relation. All of these three terms  arise from the repeated  collisions of the particle with the piston. The stationary distribution corresponding to the above Langevin equation is given by $\Psi_\text{eq}(X,V)=\mathcal{Z}^{-1}\exp{[-\beta (MV^2/2+\mathcal{U}_\text{eff}(X))]}$, where $\mathcal{U}_\text{eff}= \mathcal{U}- k_B T \ln X$ is the effective potential and $\mathcal{Z}$ is the partition function. This is consistent with the equilibrium distribution $P_\text{eq}(x,v,X,V)$ of the full system. If we replace our single particle gas by a equilibrium gas at finite density $\nu$, then  the friction coefficient  is given by   \cite{kampen1961power} $\nu \sqrt{8 m k_B T/\pi}$. Naively extending this result to our one-particle gas, by setting $\nu=1/X$, would suggest $\gamma(X)=(1/X) \sqrt{8 m k_B T/\pi}  =: \gamma_{\rm gas}$, which is different from Eq.~\eqref{gamma}. Our numerical results, however,   strongly indicates that Eq.~\eqref{eom} with $\gamma(X)$ as in Eq.~\eqref{gamma} describes the piston dynamics more  accurately. 

{\it Numerical results}.-- We compute various physical quantities related to the piston, from the exact microscopic dynamics (EMD) of particle-piston system and compare them with the corresponding results obtained from the effective Langevin equation (LE) \eqref{eom}. In the EMD simulation we start with the piston at a fixed position $X_0$ and velocity $V=0$. On the other hand, the initial position $x$ and velocity $v$  of the small particle are chosen from the equilibrium distribution $p_\text{eq}(x,v)=(1/X_0)~e^{-m v^2/(2k_B T)}/\sqrt{2 \pi k_B T/m}$ with $0<x<X_0$. Starting from this initial condition we follow the collisional dynamics. Note that there is a stochastic component to the dynamics due to the collisions between particle and heat bath. Finally we compute observables related to the piston, by averaging over the initial configurations (of particle) as well as the trajectories. In our LE simulations we start from the same fixed initial conditions $(X=X_0,V=0)$ for the piston and then average over noise realizations.  In all our simulations we have set $k_B T=1$ and $m=1$.

\begin{figure}
\centering
\includegraphics[width=3.12cm,angle=-90,keepaspectratio=true]{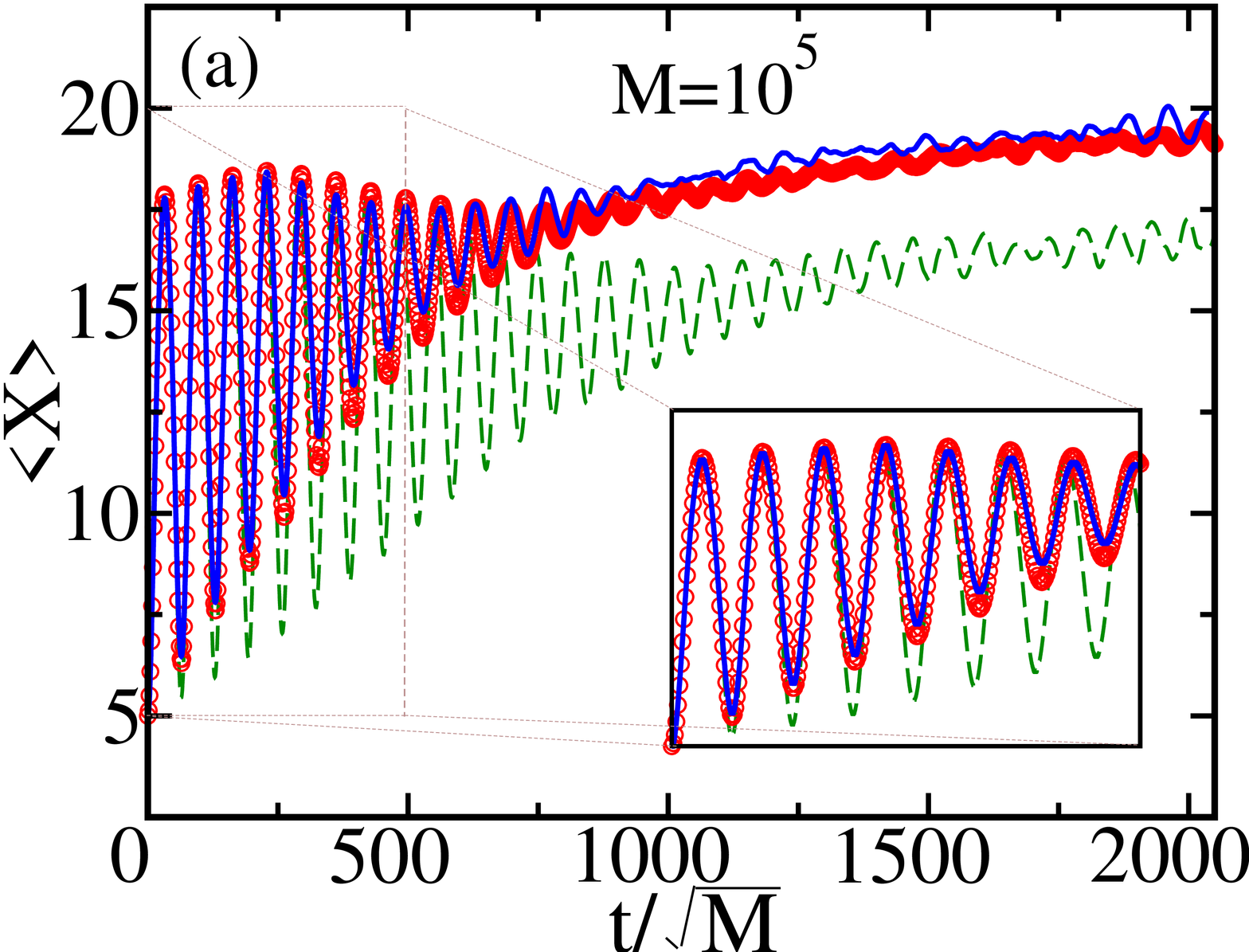}
\includegraphics[width=3.12cm,angle=-90,keepaspectratio=true]{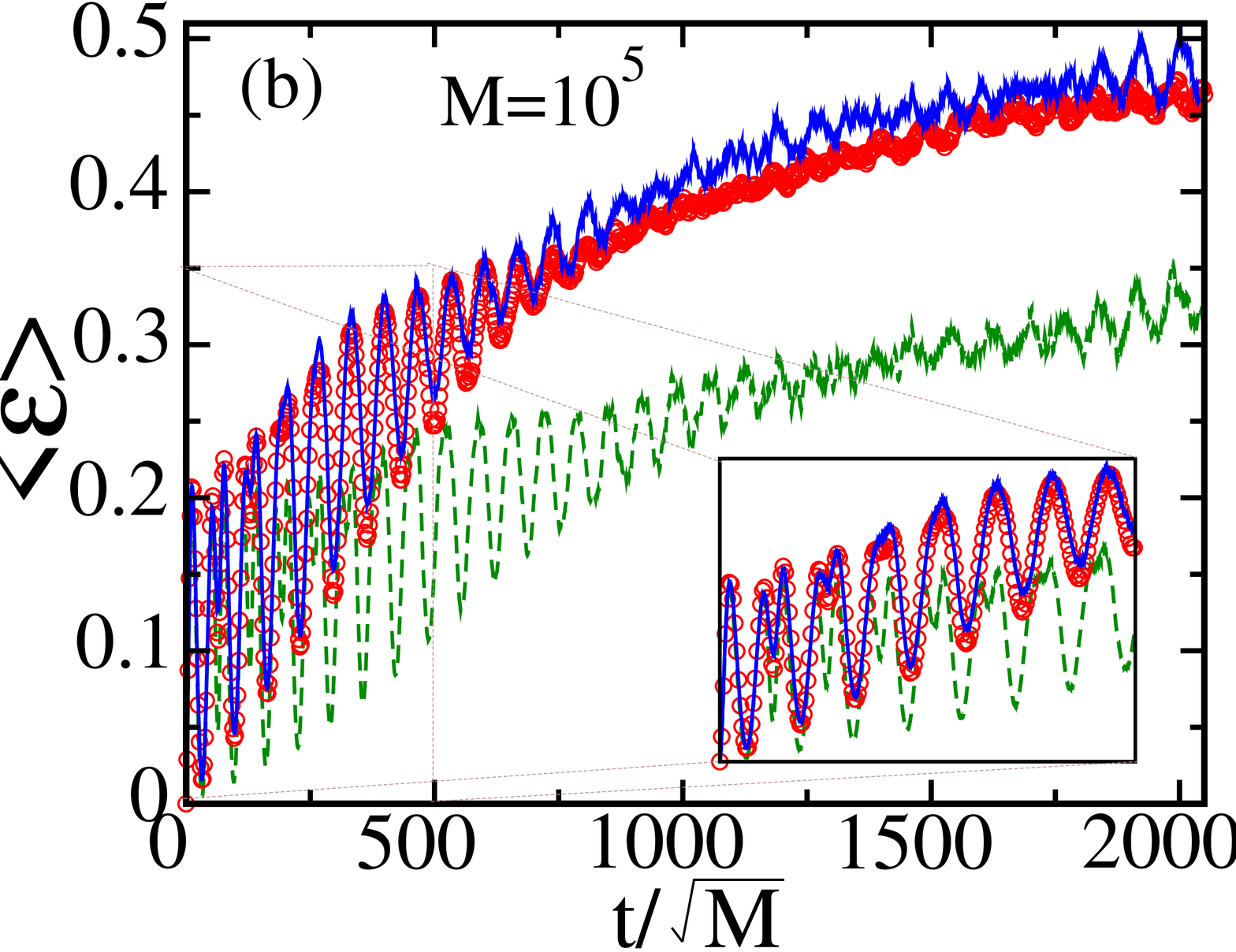}\\
\includegraphics[width=3.12cm,angle=-90,keepaspectratio=true]{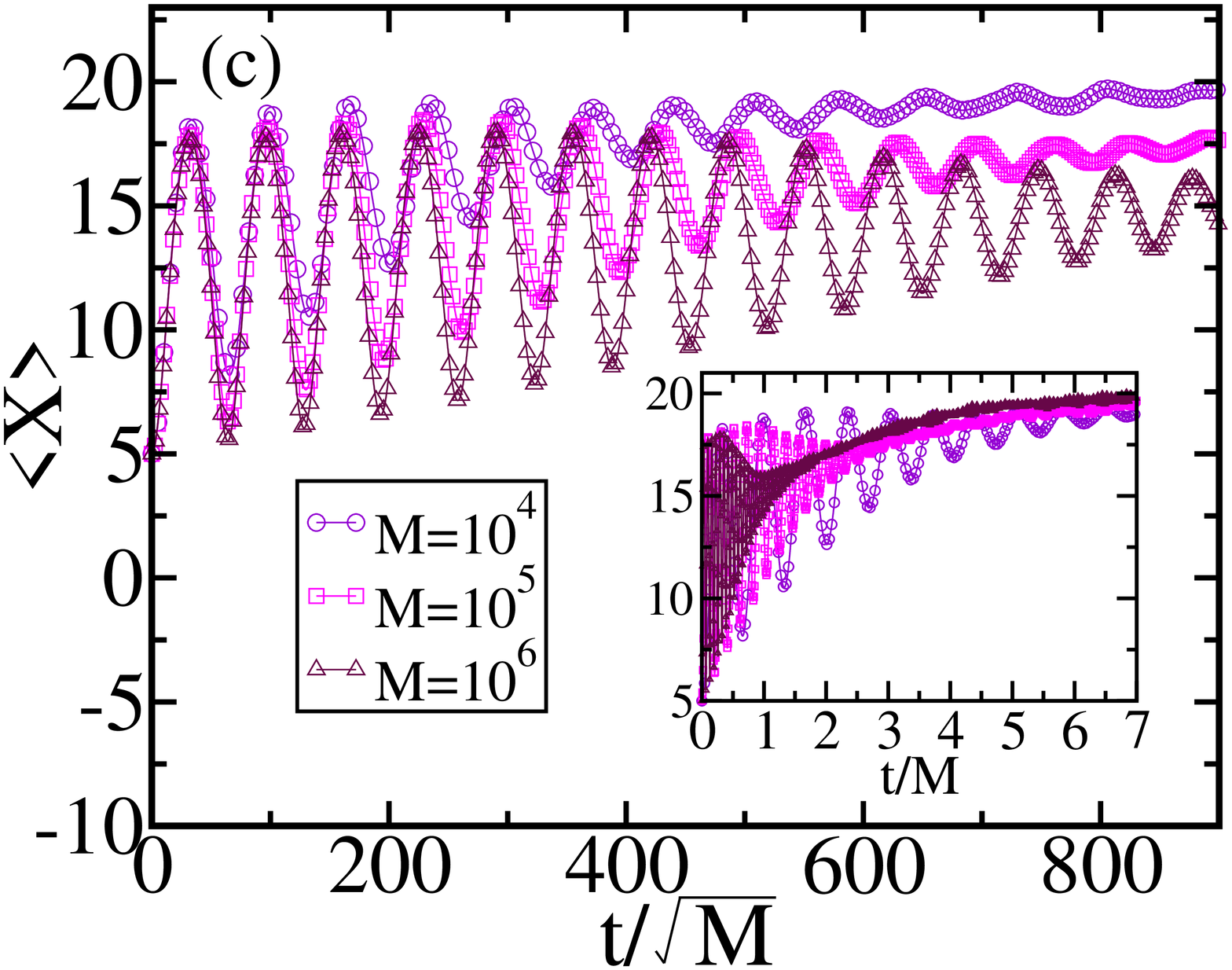}~
\includegraphics[width=3.12cm,angle=-90,keepaspectratio=true]{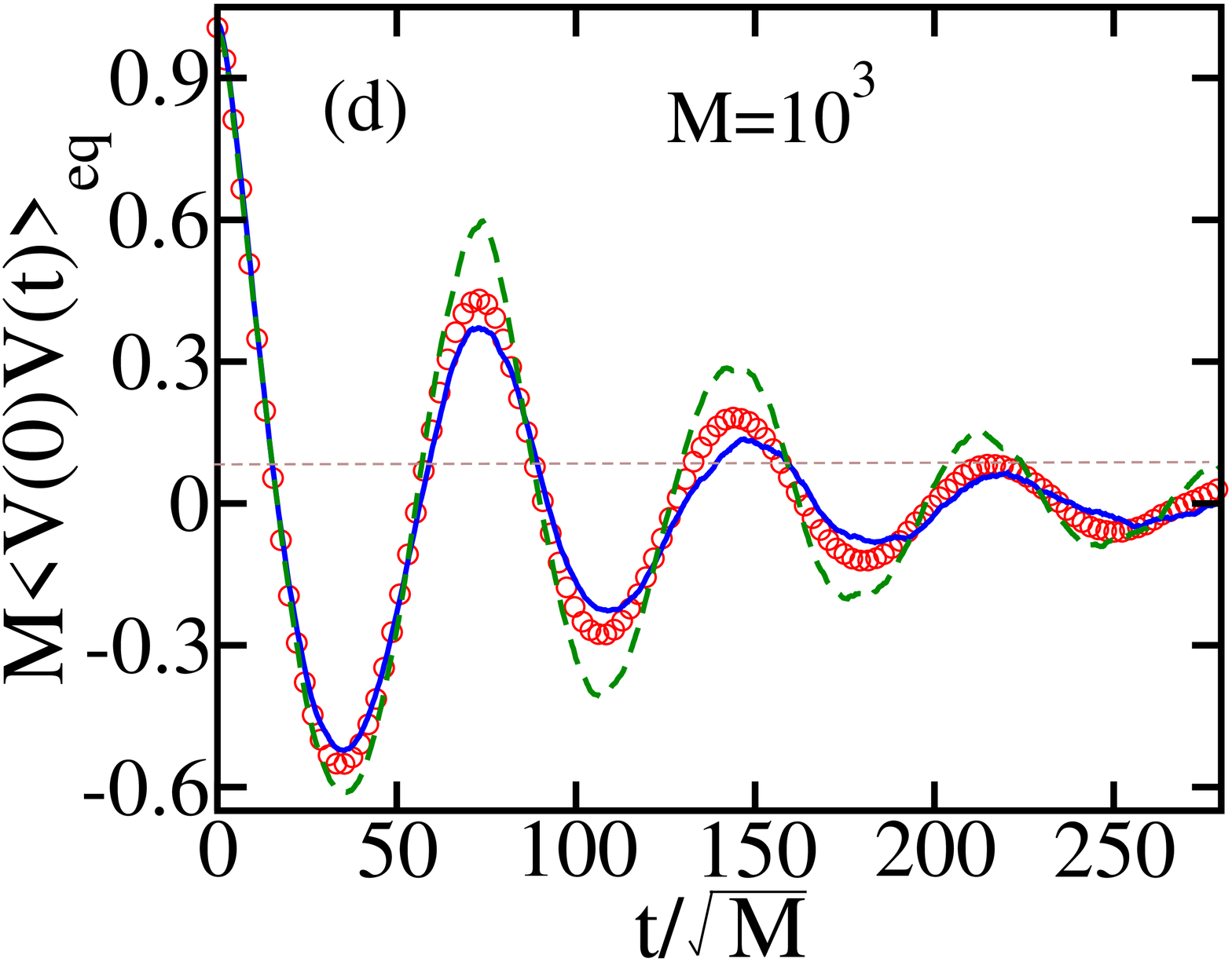}\\
  \caption{(Color online) Comparisons of data, obtained from EMD (red circles), LE with $\gamma_{\text sp}$ (blue solid line) and LE with $\gamma_{\text gas}$ (green dashed line), for  (a) average position $\langle X\ra$,  (b) kinetic energy $\la {\mathcal E} \ra$ and (d) equilibrium velocity correlation $\la V(0)V(t)\ra$ of the piston, as functions of time. The  parameter values used were $X_0=5,~F=0.1, k_B T=1$. The insets show the clear agreement between EMD and LE with $\gamma_{\text sp}$.
Figure~(c) shows that the oscillation period scales as $\sqrt{M}$ while the inset shows that the relaxation occurs at time scale of $\mathcal{O}(M)$.
}\label{f2}
\end{figure}

In Fig.~\ref{f2}(a) and (b), we plot the average position $\langle X \rangle$  and average kinetic energy $\langle \mathcal{E} \rangle$ of the piston as a function of time,  obtained from  the EMD and LE  simulations, for the case of a constant force $-F$ ($\mathcal{U}=FX,~F>0$), directed towards the bath.  We see  excellent agreement between the EMD  and the LE with $\gamma(X)$ from Eq.~\eqref{gamma}, whereas the LE with $\gamma_{\rm gas}$  shows significant deviations. We see damped oscillations and an eventual approach to the expected equilibrium values $\langle X \rangle_\text{eq}= 2k_BT/F$. Note that the mean position does not correspond to the minimum of the effective potential, $X_\text{min}=k_BT/F$,  expected from   a naive pressure balance. This is basically because of equilibration in a asymmetric potential and can also be understood in terms of a two-particle system in a constant-pressure ensemble \cite{Hondou}. From Fig.\ref{f2}(c) we can see that the time period of oscillations scales as $\sqrt{M}$ and this is again due to  the motion in the  effective potential $\mathcal{U}_{\rm eff}$. Finally the inset in Fig.~\ref{f2}(c) shows that the equilibration time scale $\sim M$, as can be inferred from the Langevin equation Eq.~\eqref{eom}.

The prediction $\gamma(X)= \gamma_{\rm gas}$ is made for a heavy particle  interacting with a many-particle gas in equilibrium. A question naturally arises whether  $\gamma(X)= \gamma_{\rm gas}$   is better than Eq.~\eqref{gamma} in describing correlations when the heavy particle is in equilibrium with the single particle gas. This leads us to investigate the velocity auto-correlation $\langle V(0)V(t) \rangle_{\rm eq}$ for the piston, as shown in Fig.~\ref{f2}(d). We plot this correlation quantity as a function of time, as obtained by simulating the EMD and from the LE with both $\gamma_{\rm sp}$ and $\gamma_{\rm gas}$. Once again we observe that $\gamma_{\rm sp}$ works much better than $\gamma_{\rm gas}$.  As  in the non-stationary case [Fig.~\ref{f2}(a) and (b)], here too the oscillation period scales as $\sqrt{M}$ while the relaxation time scales as $M$.

\begin{figure}
\centering
\includegraphics[height=2.75cm,angle=-90,keepaspectratio=true]{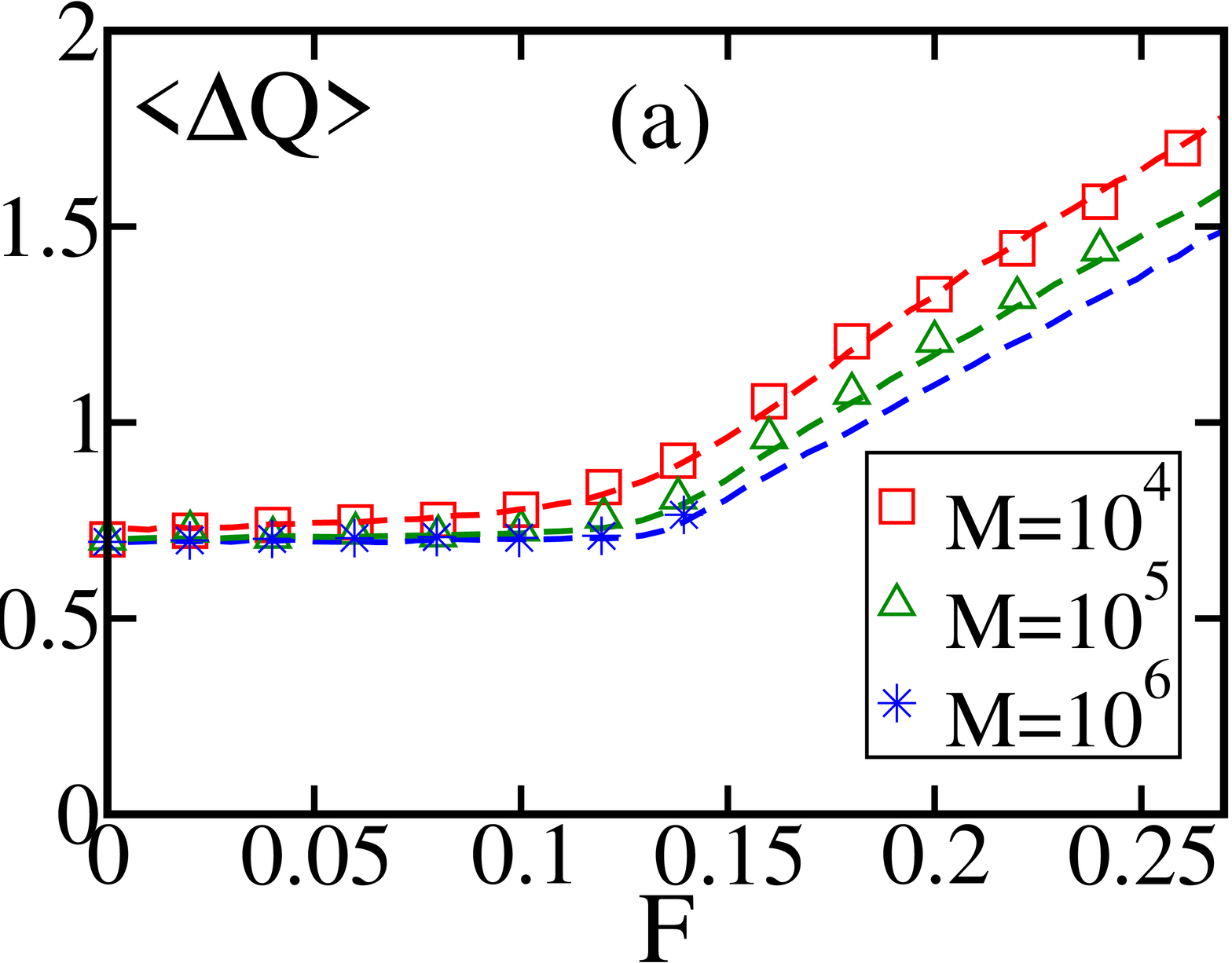}
\includegraphics[height=2.75cm,angle=-90,keepaspectratio=true]{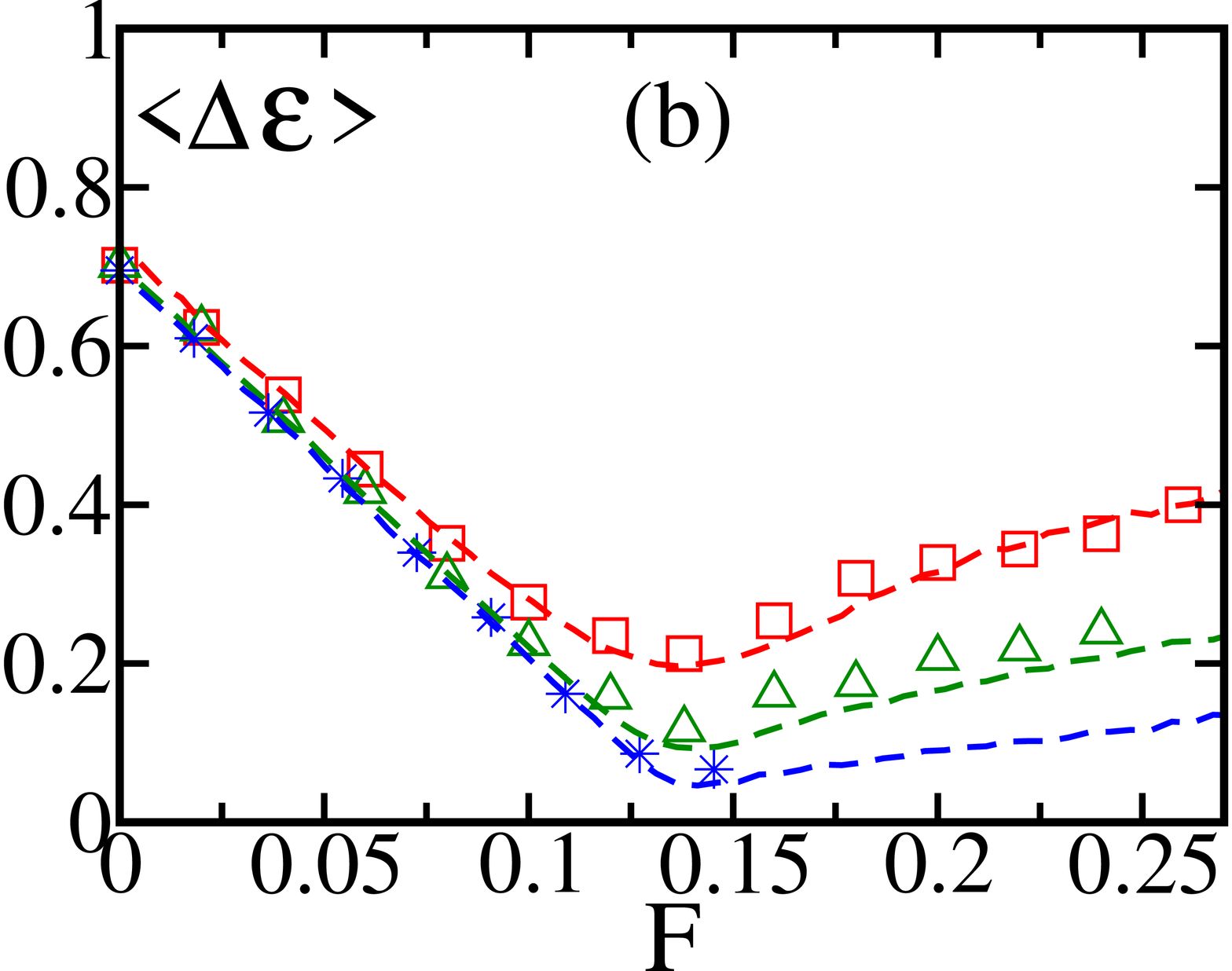} 
\includegraphics[height=2.75cm,angle=-90,keepaspectratio=true]{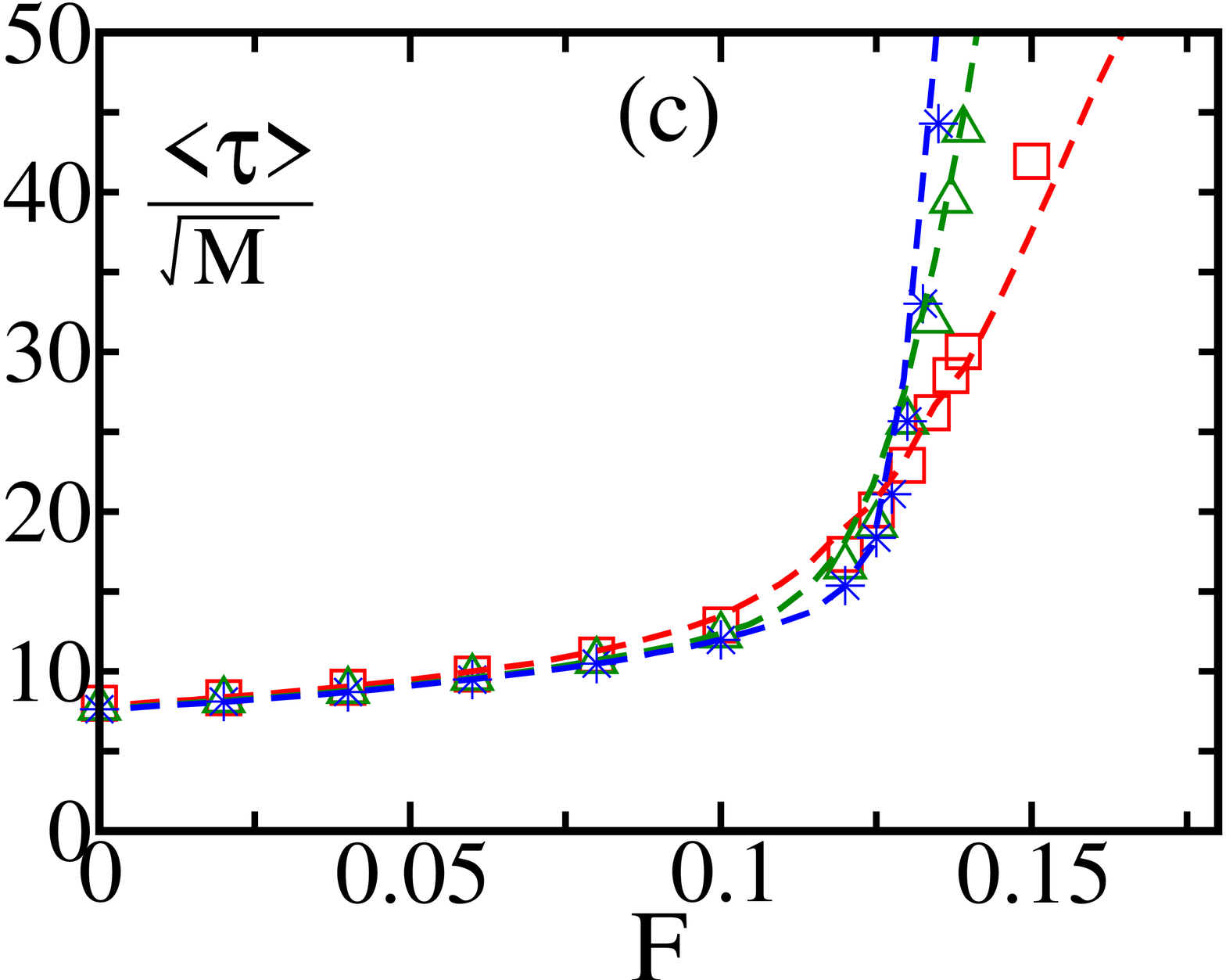}
 \caption{\label{f4}(Color online) Thermodynamic properties: (a) $\langle \Delta \mathcal{Q} \rangle$, (b) $\langle \Delta \mathcal{E} \rangle$, and (c) $\langle \tau \rangle/\sqrt{M}$ as a function of $F$, shown for different values of $M$ (same in all three plots). Symbols are from EMD simulations and dashed lines indicate the results of LE simulations (Eq.~\ref{eom} with $\gamma_{\rm sp}$ chosen from Eq.~\ref{gamma}).} \label{trajaverages}
\end{figure}

So far we have looked at averages of time-instantaneous quantities and found that the Langevin description (with $\gamma_{\rm sp}$) agrees very well with the EMD. It is natural to ask if this agreement will continue to hold for time-integrated quantities, e.g  physical observables such as heat and work. These are quantities  that are of direct relevance in the original context of the Szilard engine. We consider a protocol where the piston is released from an initial position $X_i$ and velocity $V_i$, and allowed to evolve till it reaches a specified final position $X_f$ for the first time. Note that the time $\tau$ required for this process is a random variable. For a given realization of the trajectory $\{X(t),V(t): 0 \le t \le \tau \}$,  the work done by the piston  and  the change in the kinetic energy are respectively 
\begin{eqnarray}
\mathcal{W}= \int^{X_f}_{X_i}F dX ~~{\text{and}}~~ 
\Delta \mathcal{E}=\frac{1}{2} M V_f^2-\frac{1}{2} M V_i^2~, \label{energy} 
\end{eqnarray}
where $V_f$ is the final velocity (which is  random) of the piston. From the first law of thermodynamics, the amount of heat absorbed in the process is given by
\begin{eqnarray}
 \mathcal{Q}= \Delta \mathcal{E} + \mathcal{W}~. 
\end{eqnarray}
We numerically compute these quantities from  trajectories, with $X_i=L/2, V_i=0$ and $X_f=L$,  generated by both the EMD and the LE \eqref{eom}. In Fig.~\ref{trajaverages} we plot the averages of these quantities as a function of the constant applied force $F$ and find good agreement  between the EMD and the LE. This further confirms the validity of the  Langevin description in \eqref{eom}. 

We now discuss some interesting aspects related to  Fig. \ref{trajaverages}.
In (a) where we plot $\langle \mathcal{Q} \rangle$ as a function of $F$, we see that,  at zero force and in the $M \to \infty$ limit, the amount of heat absorbed $\mathcal{Q} = \langle \Delta \mathcal{E} \rangle=k_BT \ln(2)$.  This can be understood from the LE \eqref{eom}, since we see that $\mathcal{Q}=\int_{L/2}^L dX k_B T/X +O(M^{-1/2})$, where the sub-leading correction term comes from the dissipative and noise terms. As we apply force on the piston, an amount of work $\mathcal{W}=FL/2$ is done by  the system.  Note that it is independent of trajectories and also of $M$.  On the other hand, we see  in Fig.~(\ref{trajaverages}b) that the change in average kinetic energy of the piston $\langle \Delta \mathcal{E} \rangle$ decreases with increasing $F$, as expected. In the  $M \to \infty$ limit,  $\Delta \mathcal{E}= \mathcal{Q}-\mathcal{W}=k_B T \ln (2) - FL/2$, as seen in Fig.~(\ref{trajaverages}b), for the largest mass case.   It then follows that at  a critical value of the force $F_c=2k_B T \ln 2/L$, the change in energy vanishes and all the heat absorbed gets converted to work, as in the ideal Szilard  engine. Note that in  the original single molecule Szilard engine, the work and heat computations are carried out under the assumption that the system is always in equilibrium and described by the equation of state $\langle X \rangle = k_BT/F$. However, as pointed out recently by Hondou \cite{Hondou}, in the single particle case the equation of state is in fact $\langle X \rangle = 2k_BT/F$ and this seems to contradict the basic premise of the Szilard calculation. This equation of state in fact follows on noting that $\langle 1/X \rangle = 2/\langle X \rangle$ with the average taken over $P_{\text eq}$. One of the finding of our microscopic study is that in the large piston mass limit, there is a critical  force $F_c$, when one can convert  $k_BT \ln 2$ amount of heat completely into work and this happens {\it while the system is never in equilibrium} --- hence Szilard's conclusions remain valid.

Figure (\ref{trajaverages}c) shows that the time required to reach $X_f=L$ grows rapidly beyond $F_c$, and this growth is faster for larger $M$. We note that, for $M\to \infty$, the (scaled) time  required for the piston to reach $X_f=L$, given by  $\mathcal{T}/\sqrt{M}=\sqrt{1/2}\int^{L}_{L/2} [\mathcal{U}_{eff}(L/2)-\mathcal{U}_{eff}(X)]^{-1/2}dX$, diverges at $F=F_c$. For  large but finite $M$, the piston finally reaches $X_f=L$ after a large  time because of the noise and dissipation terms. It is easy to see that the piston gains energy  $\mathcal{O}(M^{-1/2})$ from the noise term during the time period $\mathcal{O}(M^{1/2})$ required to reach $L$ [as seen in Fig.~(\ref{trajaverages}c)]. Interestingly, we see that  $\Delta \mathcal{E}$ for finite masses has a minimum at $F=F_c$ and beyond $F_c$, the $\Delta \mathcal{E}$ increases.

Qualitatively similar thermodynamic features can also be seen for other choices  of the external force, for instance harmonic force like $F=-\kappa(X-\frac{L}{2})$ or $F= -\frac{\mu}{X}$. Following the aforementioned procedure for work extraction process of Szil\'ard engine, we can show that, by suitably choosing the parameters ($\kappa=\kappa_c=8 k_BT \log(2)/L^2$ or $\mu=\mu_c=k_BT$) we can get $\langle \Delta \mathcal{E} \rangle=0$, $\langle \Delta \mathcal{Q} \rangle=k_BT\ln(2)$ and work extracted is $k_BT \ln(2)$ in the limit $\epsilon=\sqrt{m/M} \rightarrow 0$.

So far we have presented various numerical evidence which suggests that \eqref{eom} is the correct effective Langevin description of the piston motion. We now present a possible theoretical derivation based on van-Kampen's $\Omega-$expansion
method \cite{VANKAMPEN2007219}, where $\epsilon=\sqrt{m/M}$ is the expansion parameter. We start with the  master equation for the joint probability distribution function  $P(x,X,v,V,t)$ as follows~\cite{Hondou}:
\begin{eqnarray}
\frac{\partial P}{\partial t} = \mathcal{L}P~~~~;~~~~\mathcal{L}=\mathcal{L}_d+ \mathcal{L}_r+\mathcal{L}_c~. \label{me}
\end{eqnarray}
Here
\begin{eqnarray}
 \mathcal{L}_dP=-v\frac{\partial P}{\partial x} -V\frac{\partial P}{\partial X}  
+\frac{F}{M}\frac{\partial P}{\partial V} 
\end{eqnarray}
corresponds to the deterministic evolution of the particle and piston, in between collisions. The collision of the small particle with the thermal reservoir is represented by the term
\begin{align}
\begin{split}
\mathcal{L}_rP &=\int dv' \left[\mathcal{R}_r(v|v')P(x,X,v',V) \right. \\ 
& \left. 
-  \mathcal{R}_r(v'|v)P(x,X,v,V) \right]~,  
\end{split} \label{collwall} 
\end{align}
with  $\mathcal{R}_r(v|v')=\delta(x) (-v') ~\theta(-v')f(v)$. The first term in \eqref{collwall} corresponds to gain in probability from events in which a particle with a negative velocity hits the reservoir and emerges with a velocity $v$ with probability $f(v)$. The second term corresponds to loss of probability when a particle with velocity $v$ hits the reservoir. Finally the elastic collisions between the particle and piston are represented by the term 
\begin{align}
\begin{split}
 \mathcal{L}_cP &= \int dv' dV' \left[\mathcal{R}_c(v,V |v',V')P(x,X,v',V') \right. \\ &\left. - \mathcal{R}_c(v',V'|v,V)P(x,X,v,V) \right],~
\end{split} \label{collpiston}
\end{align}
with $ \mathcal{R}_c(v,V|v',V')=  \delta(X-x)\theta(v'-V') (v'-V') \delta\left[v'-\frac{(\epsilon^2-1)v+2V}{1+\epsilon^2}\right] \delta\left[V'-\frac{2\epsilon^2 v+(1-\epsilon^2)V}{1+\epsilon^2} \right] $. The $\delta-$function constraints on the velocities arise from the momentum and energy conservation during the elastic collisions. Our aim now is to integrate out the particle-degrees of freedom, to get an effective Fokker-Planck equation for the piston.

We first rescale the velocity of the piston and the particle as $U=V\sqrt{\beta M}$, $u=v\sqrt{\beta m}$ and time as $\tau=t/\sqrt{\beta m}$. We write the joint distribution of the rescaled variables $Q(x,X,u,U,\tau)$  as 
\begin{eqnarray}
 Q(x,X,u,U,\tau)=\Phi(x,u,\tau|X,U)\Psi(X,U,\tau), \label{Q}
\end{eqnarray}
where $\Psi(X,U,\tau)$ is the marginal distribution of the position and velocity of the piston and $\Phi(x,u,\tau|X,U)$ is the distribution of the position and velocity of the small particle conditioned on given piston configuration. Inserting this form of $Q(x,X,u,U,\tau)$ in \eqref{me}, and performing some simplifications we obtain a master equation 
\begin{eqnarray}
\frac{\partial Q}{\partial \tau}=&&\mathcal{L}_\epsilon Q \label{Q-me}
\end{eqnarray}
where the expansion parameter $\epsilon$ is explicit \cite{SM}.
To proceed further we expand the operator $\mathcal{L}$ and the distributions  $\Phi$, $\Psi$ in powers of $\epsilon$ as follows \redw{\cite{SM}}:
\begin{eqnarray}
\mathcal{L}&=\mathcal{L}_0+\epsilon \mathcal{L}_1 + \epsilon^2 \mathcal{L}_2 +\mathcal{O}(\epsilon^3), \label{FP-O}\\
\Phi&=\Phi_0+\epsilon \Phi_1 + \epsilon^2 \Phi_2 + \mathcal{O}(\epsilon^3), \label{phi}\\
\Psi&=\Psi_0+\epsilon \Psi_1 + \epsilon^2 \Psi_2 +\mathcal{O}(\epsilon^3). \label{psi}
\end{eqnarray}
Inserting  these  in the master equation \eqref{Q-me}, we  look at the resulting equation at different orders of $\epsilon$. At each order we integrate the particle position $x$ and velocity $u$ to get following time evolution equations for $\Psi_0,~\Psi_1$ and~$\Psi_2$. 
\begin{align}
&\frac{\partial \Psi_0}{\partial \tau}=0 \nonumber  \\
&\frac{\partial \Psi_1}{\partial\tau}   + U\frac{\partial \Psi_0}{\partial X}  - \beta  F\frac{\partial \Psi_0}{\partial 
U} =-\frac{1}{X}  \frac{\partial \Psi_0}{\partial U} \\
&\frac{\partial \Psi_2}{\partial\tau}   + U\frac{\partial \Psi_1}{\partial X}  + \left[-\beta  F 
+\frac{1}{X}\right]\frac{\partial \Psi_1}{\partial U}= \gamma(X)\frac{\partial }{\partial U}\left[U\Psi_0 + 
\frac{\partial \Psi_0}{\partial U} \right]  \nonumber
\label{psinotdot}
\end{align}
which provide the equation for $\Psi = \Psi_0 + \epsilon \Psi_1 + \epsilon^2 \Psi_2 + \mathcal{O}(\epsilon^3)$. In the original variables, we find 
\begin{eqnarray}
\frac{\partial \Psi}{\partial t} &=&  - V\frac{\partial \Psi}{\partial X}  - \frac{1}{M} \left(-F+\frac{k_BT}{X}\right)\frac{\partial 
\Psi}{\partial V}   \nonumber \\
&+& \frac{\gamma(X,t)}{M} \frac{\partial  \left(V\Psi \right) }{\partial V} +\frac{\gamma(X,t) k_BT}{M^2}  \frac{\partial^2 \Psi}{\partial V^2} 
\label{psi-FP}
\end{eqnarray}
where the expression of $\gamma(X,t)$ is given in terms of an inverse Laplace transform in \redw{\cite{SM}}. The small and large time asymptotic forms of $\gamma(X,t)$ are 
\begin{eqnarray}
\gamma(X,t) ~\sim
\begin{cases}
& \frac{1}{X}\sqrt{\frac{8 mk_BT}{\pi}} ~~~;~~~t\rightarrow 0~~\nonumber\\
& \nonumber \\
& \frac{\sqrt{mk_BT}}{X}\log(t) ~~~;~~~t \rightarrow \infty.
\end{cases}
\label{gamma-p}
\end{eqnarray}
Note that the Fokker-Planck Eq.~\eqref{psi-FP} corresponds precisely to the Langevin equation \eqref{eom}. However, our previously presented numerical evidences suggest that the friction coefficient is given by Eq.~\eqref{gamma} which is different from the prediction in Eq. \eqref{gamma-p}.

At short times the predicted $\gamma(X)$ is same as $\gamma_{\text gas}$ while at large times it diverges logarithmically. This divergence indicates the breakdown of the perturbation theory, which however provides the correct form of equation of motion. The possible reasons for this breakdown are the following:  (i) The perturbation theory here implicitly assumes a separation of time scales. As pointed out in a recent paper, for the case of a fixed piston the small particle shows a slow power-law relaxation to equilibrium \cite{Bhat}. This suggests that there may be no time scale separation between the particle and the piston. (ii) In addition, we have not taken the multiple collisions into account. These two means of breakdown arise because in the perturbation expansions in Eqs. \eqref{FP-O}-\eqref{psi}, we have implicitly assumed that the velocity $v$ of the small particle is always of order $\sqrt{k_BT}$. This is not always true because it is possible that the small particle can emerge from the bath with very small value ($\sim \epsilon$) and then the above perturbation expansion fails.
Hence the derivation of exact expression for $\gamma$ is non-trivial.  More details about the derivation of Eq.\ref{psi} are given in the supplementary material.

{\it Conclusion:} In this Letter we looked at the non-equilibrium dynamics of the Szilard engine. We find that in the limit of large mass of the piston, it's effective stochastic dynamics is given by  a Langevin  equation (Eq.~\ref{eom}) with a space dependent friction coefficient $\gamma(X)$ (Eq.~\ref{gamma}).  To arrive at this equation we integrated out the particle degrees of freedom in the full master equation following the $\Omega$-expansion method. While this perturbation method correctly provides the form of the Langevin equation, it does not give the right form for the friction term, and we argue that this arises due to rare events in the small particle's dynamics.  However our extensive numerical studies suggests that the form of the friction coefficient $\gamma(X)$ given in Eq. \eqref{gamma} is in fact accurate. To verify this form further, we have also considered the situation where the piston interacts with one thermal particle on each side. In this case also we find excellent numerical agreement  (see \cite{SM}).  Finally we used  the Langevin equation description to study the thermodynamics of work extraction process in the Szilard engine. We found that, in the limit $M\rightarrow \infty$, and at a crucial value of the force $F_c$, the work extracted from the engine  becomes equal to the heat absorbed, $k_BT \ln (2)$   --- even though the piston remains out of equilibrium during the entire process. 
The analytic derivation of the result $\gamma(X)=\gamma_{sp}=(1/X)\sqrt{8\pi mk_BT}$ for the dissipation constant remains an interesting open problem.


\section{Acknowledgements}
D.B. acknowledges Arghya Dutta, Udo Seifert, Onuttom Narayan, Tridibh Sadhu and Satya Majumdar for discussions. A.D., A.K. and S.S. acknowledge support of the Indo-French Centre for the promotion of advanced research (IFCPAR) under Project No. 5604-2. A. K. acknowledges support from DST grant under project No. ECR/2017/000634.
\bibliographystyle{apsrev4-1}
\bibliography{reference}

\end{document}